\documentstyle[12pt,supercite]{article}

\newcommand{\Gdx}{\mathop{G^{^{\rm diam}}(x)}}
\newcommand{\Gd}{\mathop{G^{^{\rm diam}}}}
\newcommand{\gx}{\mathop{g^{(0)}(x)}}
\newcommand{\g}{\mathop{g^{(0)}}}

\newcommand{\AmS}{{\protect\the\textfont2
  A\kern-.1667em\lower.5ex\hbox{M}\kern-.125emS}}

\def\abstracts#1#2#3{{
        \centering{\begin{minipage}{4.62in}\baselineskip=13pt
        \small
        \centerline{\bf Abstract}
        \vspace*{0.2cm}                % W. Janke (July 1, 1992)
        \parindent=0pt #1\par
        \parindent=18pt #2\par
        \parindent=15pt #3
        \end{minipage} }\par}}
\begin{document}
\vspace*{-2cm}
\hfill \parbox{5cm}{ KOMA-96-33   \\
                     July 1996} \\
\vspace*{2cm}
\centerline{\Large \bf 
                       Monte Carlo Study of Cluster-Diameter Distribution:
           }\\[0.3cm]
\centerline{\Large \bf 
                       A New Observable to Estimate Correlation Lengths
           }\\[0.8cm]
\vspace*{0.2cm}
\centerline{\large {\em Wolfhard Janke\/} and
                   {\em Stefan Kappler\/}}\\[0.4cm]
\centerline{\large {\small Institut f\"ur Physik,
                    Johannes Gutenberg-Universit\"at Mainz}}
\centerline{    {\small Staudinger Weg 7, 55099 Mainz, Germany }}\\[0.5cm]
\abstracts{}{
We report numerical simulations of two-dimensional $q$-state Potts models
with emphasis on a new quantity for the computation of spatial correlation 
lengths. This quantity is the cluster-diameter distribution function 
$G_{\rm diam}(x)$, which measures the distribution of the diameter of 
stochastically defined cluster. Theoretically it is predicted to fall off 
exponentially for large diameter $x$, $G_{\rm diam} \propto \exp(-x/\xi)$,
where $\xi$ is the correlation length as usually defined through the 
large-distance behavior of two-point correlation functions. The results of 
our extensive Monte Carlo study in the disordered phase of the
models with $q=10$, 15, and $20$ on large square lattices of size 
$300 \times 300$, $120 \times 120$, and $80 \times 80$, respectively, clearly
confirm the theoretically predicted behavior. Moreover, using this observable
we are able to verify an exact formula for the correlation length 
$\xi_d(\beta_t)$ in the disordered phase at the first-order transition 
point $\beta_t$ with an accuracy of about $1\%-2\%$
for all considered values of $q$. This is a considerable improvement over 
estimates derived from the large-distance behavior of standard
(projected) two-point correlation functions, which are also discussed for
comparison.\\[2cm]
}{}
%\vspace*{0.5cm}
\thispagestyle{empty}
\newpage
\pagenumbering{arabic}
%
%---------------------------------------------------------
                     \section{Introduction}
%---------------------------------------------------------
%
The physics of phase transitions is essentially governed by the behavior of
the spatial correlation length $\xi$. While in some problems, e.g. at a 
continuous phase transition where $\xi$ diverges, it is often sufficient to
know the qualitative behavior, there are also many applications which rely 
on quantitative estimates of $\xi$. This applies in particular to the 
finite-size scaling behavior near a first-order phase transition \cite{review}
where $\xi$ stays finite and sets the length scale above which asymptotic 
considerations should apply \cite{athens}. Since analytical predictions are 
scarce it is therefore of great practical importance to develop refined 
numerical methods for reliable computations of correlation lengths.

In order to evaluate the accuracy of a newly proposed method one should apply
it first to models where analytical predictions are available. The best known 
example is the two-dimensional (2D) Ising model where $\xi$ is exactly known
at all temperatures \cite{ising}. But already the generalization to the 2D 
$q$-state Potts model \cite{wu} complicates the theoretical analysis 
considerably, and much less is known analytically. It was therefore a great
success when a few years ago at least the correlation length $\xi_d(\beta_t)$ 
in the disordered phase {\em at\/} the first-order transition point $\beta_t$
for $q \ge 5$ could be calculated exactly \cite{xi_1,xi_2,bj92}. Apart from 
heuristic arguments, no analytical predictions are available for the 
correlation length $\xi_o(\beta_t)$ in the ordered phase, and previous 
numerical simulations \cite{previous1,previous2,previous3} turned out to be 
difficult to interpret. This was
the physical motivation to start a project \cite{jk95a,jk95b} with the goal 
to clarify conflicting conjectures for the ratio $\xi_o/\xi_d$ at $\beta_t$.
The idea was, of course, to test the employed numerical methods first for 
the exactly known correlation length in the disordered phase \cite{jk95a} 
and then to proceed to the so far unexplored ordered phase \cite{jk95b}.

One often employed way to extract correlation lengths is to study the 
exponential decay of two-point correlation functions in the asymptotic limit
of large distances. While this methods works perfectly for the 2D Ising model,
for $\xi_d(\beta_t)$ of the 2D $q$-state Potts models with $q=10$, 15, and 20
we experienced quite nasty systematic deviations from the exact answer by 
about $10\% - 20\%$ \cite{jk95a}. The deviations could be traced back to the
unexpected importance of higher order excitations, but even though the 
Monte Carlo simulations were performed on quite large lattices and with a 
high statistics of about $50\,000 - 100\,000$ uncorrelated measurements, 
least-square fits with sufficiently many correction terms turned out to be 
too unstable to predict reliable numbers.

A way out of this problem is to search for a different estimator of $\xi$
which is less affected by correction terms. A systematic search is certainly
very difficult, but one possible candidate was recently suggested in 
analytical work \cite{bc94} making extensively use of the Fortuin-Kasteleyn
cluster representation \cite{FoKa} of the Potts model. In Ref.~\cite{bc94} it
was shown that the distribution of the cluster diameter, $G_{\rm diam}(x)$,
decays exponentially for large diameter $x$, and that the decay constant is 
identical to the inverse correlation length (as defined from the decay of the
two-point correlation function). This prompted us to investigate if the 
cluster-diameter distribution function is better suited for a numerical 
determination of the correlation length. In the following we report 
high-statistics Monte Carlo simulations of the models with $q=10$, 15, and 20,
focussing on the properties of the new observable. As the main
result it turns out to be indeed very well suited in the disordered phase, 
allowing for the first time a confirmation of the analytical formula for 
$\xi_d(\beta_t)$ with an accuracy of about $1\% - 2\%$. Since we used larger
lattices and considerably higher statistics than in our previous 
studies \cite{jk95a}, we discuss for comparison also the newly obtained 
estimates for $\xi_d(\beta_t)$ from two different projections of the 
standard two-point correlation function.  

The remainder of the paper is organized as follows. In Sec.~2 we first recall
the definition of the model and some exact results. We then discuss the 
simulation techniques and in particular describe the various estimators used 
to measure the correlation length. The results of our simulations are 
presented in Sec.~3, and in Sec.~4 we conclude with a brief summary of the
main results and some final remarks.
%
%---------------------------------------------------------
           \section{Model and observables}
%---------------------------------------------------------
%
In our Monte Carlo simulations we used the standard definition of the Potts
model partition function \cite{wu},
\begin{equation}
  Z = \sum_{\{s_i\}} e^{-\beta E}; \, E = -\sum_{\langle ij \rangle}
  \delta_{s_i s_j}; \, s_i = 1,\dots,q,
\label{eq:model}
\end{equation}
where $\beta = J/k_BT$ is the inverse temperature in natural units, $i$ denote
the lattice sites of a square lattice, $\langle ij \rangle$
are nearest-neighbor pairs, $\delta_{s_i s_j}$ is the Kronecker delta 
symbol, and $q$ is the number of states per spin. In all simulations we used
periodic boundary conditions to minimize finite-size effects.

In the following we report results for the models with $q=10$, 15, and 20,
employing lattices of size $V =L \times L$ with $L = 300$, 120, and 80,
respectively. All simulations were performed in the canonical ensemble at 
the infinite volume first-order transition point 
$\beta_t = \ln (1+\sqrt{q})$, at which the ordered and disordered phase can
coexist. In a Monte Carlo simulation, the system can be biased into one 
of the two phases by the choice of the initial spin configuration.  
To update the spins we used the Wolff single-cluster
algorithm \cite{wolff_algo}. From a previous comparative study \cite{jk95a}
we knew that in the disordered phase this
algorithm clearly outperforms all other standard algorithms such as Metropolis,
heat-bath, and Swendsen-Wang multiple cluster \cite{sw}.

The lattice sizes were chosen such that, for each value of $q$, 
$L \approx 28 \xi_d(\beta_t)$, with $\xi_d(\beta_t) = 10.559519\dots$,
$4.180954\dots$, and $2.695502\dots$ for $q=10$, 15, and 20, 
respectively \cite{xi_1,xi_2,bj92}. Starting from a completely random 
configuration of spins it is then extremely probable
that the system will stay in the disordered phase for a sufficiently long
time, allowing statistically meaningful measurements of quantities being
characteristic for the pure disordered phase. More precisely, by recalling
that the escape probability ${\cal P}$ is proportional to 
$\exp(-2\sigma_{od}L)$ and that in two dimensions the interface tension 
$\sigma_{od}$ can be expressed in terms of the correlation length of the 
disordered phase \cite{bj92}, $\sigma_{od} = 1/2\xi_d(\beta_t)$, one easily
arrives at the order-of-magnitude estimate 
${\cal P} \propto \exp(-L/\xi_d(\beta_t)) \approx \exp(-28) \approx 10^{-12}$.
Finite-size corrections in the pure disordered phase are expected to be of 
the same order.

In this work we mainly focussed on measurements of the
probability distribution of the cluster diameter, ${\rm diam} \,C_{i_0}$,
which, in general, is defined as the ma\-xi\-mal extension of a cluster 
in any of the $D$ coordinate directions of a hypercubic lattice; for an 
illustration see Fig.~\ref{fig:sketch}. The cluster-diameter distribution
function $\Gdx$ is then the probability,
\begin{equation}
\Gdx = \mu(\, {\rm diam} \,C_{i_0} = x),
\label{eq:gdiam}
\end{equation}
that the cluster $C_{i_0}$ connected to a lattice site $i_0$ has a given
diameter $x$ \cite{bc94}. To increase the statistics we took advantage of the
periodic boundary conditions and ave\-raged $\Gdx$ over all lattice sites
$i_0$. In practice this amounts to recording a histogram $H^{\rm diam}(x)$,
whose entries at $x= {\rm diam} \,C$ are incremented by the size or weight
$|C|$ of each simulated cluster. If properly normalized, $H^{\rm diam}(x)$
is then an estimator of the probability distribution $\Gdx$. As discussed in
the Introduction the theoretically expected asymptotic behavior of $\Gdx$ in
the disordered phase is an exponential decay governed by the correlation 
length $\xi_d$ \cite{bc94}, 
\begin{equation}
\Gdx = a \exp(-x/\xi_d) + \dots.
\label{eq:Gdfit}
\end{equation}
By taking the logarithm of $\Gd$ and performing linear two-parameter fits
it is then straightforward to extract $\xi_d$.

For comparison we considered also in the new simulations the
$k_y^{(n)} = 2\pi n/L$ momentum projections ($i=(i_x,i_y)$),
\begin{equation}
  g^{(n)}(i_x,j_x) = \frac{1}{L} \sum_{i_y,j_y} G(i,j)
  e^{i k_y^{(n)}(i_y-j_y)},
\label{eq:g}
\end{equation}
with $n=0$ and 1 of the two-point correlation function
\begin{equation}
  G(i,j) \equiv \langle \delta_{s_i s_j} - \frac{1}{q} \rangle.
\label{eq:G}
\end{equation}
For the measurements we actually decomposed the whole spin configuration
into stochastic Swendsen-Wang (multiple) cluster and used the improved
cluster estimator \cite{sokal}
\begin{equation}
  G(i,j) = \frac{q-1}{q} \langle \Theta(i,j) \rangle,
\label{eq:Gimp}
\end{equation}
where $\Theta(i,j)=1$, if $i$ and $j$ belong to the same cluster, and
$\Theta = 0$ otherwise. In particular for small average single-cluster sizes 
(cp. Table~\ref{tab:stat}), this procedure is more efficient than using
directly the corresponding improved single-cluster estimator.

As discussed previously \cite{jk95a}, to extract $\xi_d$ from the 
large distance behavior of (\ref{eq:g}),
non-linear four-parameter fits of the form
\begin{equation}
  g^{(n)}(x) \equiv g^{(n)}(i_x,0) = a \,{\rm ch}(   \frac{L/2-x}{\xi_d^{(n)}} )
  +   b \,{\rm ch}( c \frac{L/2-x}{\xi_d^{(n)}} ),
\label{eq:fit_4}
\end{equation}
with
\begin{equation}
\xi_d \approx \xi_d^{(n)} / \sqrt{1 \!-\! (2 \pi n \xi_d^{(n)}/L)^2}.
\label{eq:xi}
\end{equation}
are necessary. Below we shall report results for the first two projections
with $n=0$ and $n=1$. While the $n=0$ projection has been studied before on
smaller lattices \cite{jk95a}, the use of the $n=1$ projection
in the disordered phase is actually also new. Originally this projection
was applied in the ordered phase where it is essential for removing
constant background terms caused by the non-zero 
magnetization \cite{previous3,jk95b}. Notice that for the large lattice sizes
used in this study, $L \approx 28 \xi_d$, the difference in (\ref{eq:xi}) 
between the fit parameter $\xi_d^{(n)}$ and $\xi_d$ is only about $2.4\%$.

The computer code was implemented on a T3D parallel computer in a trivial
way by running 64 independent simulations in parallel.
This allowed us to generate the very high statistics
compiled in Table~\ref{tab:stat}. Here we followed the usual convention
and defined $V/\langle |C| \rangle_{\rm SC}$ single-cluster steps as one
Monte Carlo update sweep (MCS), where $\langle |C| \rangle_{\rm SC}$
is the average cluster size, and rescaled the integrated autocorrelation
time of the internal energy, $\tau_{\rm int,e}$, to this unit of time.
Per $\tau_{\rm int,e}$ we performed about two measurements of the projected
correlation functions. The size and diameter of the clusters were measured
for each generated cluster. The statistical error bars are estimated from 
the fluctuations among the 64 independent copies by using the standard 
jack-knife procedure \cite{jack}. The total running time of the simulations
amounts to about five years of CPU time on a typical workstation.
%
%---------------------------------------------------------
           \section{Results}
%---------------------------------------------------------
%
In all simulations we monitored the time evolution of the energy and 
magnetization to convince ourselves that the system never escaped into 
the ordered phase. As a more quantitative measure we also computed energy
and magnetization moments which can be compared with exact \cite{wu,baxter1}
or series expansion results \cite{large_q}. The average and maximum cluster
sizes and the maximum cluster diameter found in the 
simulations are given in Table~{\ref{tab:stat}. As a result
of these tests we are convinced that, despite the very long
run times, our results for $\xi_d$ can be identified with the {\em pure} 
disordered phase correlation length.

The data for $\Gdx$ and $\gx$ are shown for $q=10$, 15, and 20 in the 
semi-log plots of Fig.~\ref{fig:gdiam}. The continuous lines are 
one- and three-parameter fits to the Ansatz (\ref{eq:Gdfit}) and 
(\ref{eq:fit_4}), respectively, with $\xi_d$ held fixed at its theoretical
value ($ = 10.559519\dots$, $4.180954\dots$, and $2.695502\dots$ for $q=10$,
15, and 20 \cite{xi_1,xi_2,bj92}). Let us first concentrate on the new 
observable, the cluster-diameter probability distribution $\Gdx$. At first
sight the constrained one-parameter fit to $\Gd$ looks
less perfect than the constrained three-parameter fit to $\g$, since the data
points are more randomly scattered around the fit. The reason is that the
correlations between the estimates at $x$ and $x + \Delta x$ are much smaller
for $\Gdx$ than for $\gx$. This can be understood by noting that a cluster of
diameter $x_0$ contributes only to the {\em one\/} estimate
of $\Gdx$ at $x=x_0$, but to {\em all\/} estimates of $\gx$ with $x \le x_0$
(recall the cluster estimator (\ref{eq:Gimp})).

The correlation length estimates resulting from various unconstrained
two-pa\-ra\-me\-ter fits to $\Gd$ in intervals $x_{\rm min} \dots x_{\rm max}$
with $x_{\rm max} = 130$, 50, and 40 for $q=10$, 15, and 20, respectively,
are collected in Table~\ref{tab:xi.2Lx2L.fit}. 
We see that the results are in very good agreement with the theoretically
expected values, with only slight systematic deviations of about $1\%-2\%$.
Contrary to the results \cite{jk95a} obtained from $\gx$ the fitted values 
tend now to be overestimates for small $x_{\rm min}$. This tendency becomes
obvious in Fig.~\ref{fig:gdiam_eff} where we show the effective correlation 
lengths 
\begin{equation}
\xi^{\rm eff}_d(x) = 1/\ln [C(x)/C(x+1)],
\label{eq:xi_eff}
\end{equation}
with $C = \Gd$ or $\g$. The $\xi^{\rm eff}_d(x)$ are just the inverse of the
local slopes in Fig.~\ref{fig:gdiam}. By recalling that neighboring values of
$\Gd$ are much less correlated than those of $\g$, this explains
the much larger error bars on the data for $\xi^{\rm eff}_d$ derived
from $\Gd$. Observe that $\xi^{\rm eff}_d$ obtained from $\Gd$
develop a much more pronounced plateau for $q=15$ and $20$ than for $q=10$,
before also here the statistical errors increase
and the data start to fluctuate around the theoretically expected value.
To conclude this subsection, by using the cluster-diameter probability
distribution as an estimator for the correlation length, we succeeded to
confirm the theoretical prediction for $\xi_d(\beta_t)$ at a $1\% - 2\%$
level.

It would of course be unfair to compare the final estimates obtained from
$\Gdx$ of the present study with the results from $\gx$ of previous 
work \cite{jk95a} which used smaller lattices and lower statistics. 
In the present study we have therefore analyzed again $\gx$. Furthermore we
discuss for the first time also $g^{(1)}(x)$ in the disordered phase. In 
Fig.~\ref{fig:gdiam} we see that the constrained three-parameter fit to $\g$
yields an excellent description of the fall-off of $\gx$ over more than four
decades. Still, from an unconstrained four-parameter fit over the same 
$x$ range with $\xi_d$ as a free parameter we obtain for $q=10$ an about 
$10\%$ smaller value of $\xi_d = 9.5(4)$. This confirms our earlier 
observation in Ref.~\cite{jk95a} that four-parameter fits
to $\g$ systematically underestimate $\xi_d$. This is demonstrated 
in more detail in Table~\ref{tab:xi.2Lx2L.fit} where we have collected
the results of various fits in the intervals $x_{\rm min} \dots x_{\rm max} =
L/2$. For all three values of $q$ we observe a clear tendency of increasing 
estimates for $\xi_d$ with increasing $x_{\rm min}$. Still, the estimates in
the last line for $\gx$ are about $8\%$ - $10\%$ below the theoretical
values. This tendency is also clearly visible in the behavior of the
$\xi^{\rm eff}_d(x)$ of $\g$ shown in Fig.~\ref{fig:gdiam_eff}.
Compared with Ref.~\cite{jk95a}
the statistics of the present simulations is higher by more than one order of
magnitude. This allowed us to include larger $x$ values in the fits and,
as expected, improved the estimates of $\xi_d$, in particular for
$q=15$ and 20. This clearly indicates that
by further increasing the statistics also the remaining discrepancies could
be removed. We can therefore conclude that there is nothing wrong,
in principle, in using the standard two-point correlation function to
estimate $\xi_d$. Numerically, however, accurate estimates would require 
an enormous effort and would thus be a rather expensive enterprise.

In Table~\ref{tab:xi.2Lx2L.fit} we also give the results of unconstrained 
four-parameter fits of the form (\ref{eq:fit_4}) to 
$g^{(1)}(x)$ where, by using (\ref{eq:xi}), we have already converted 
$\xi_d^{(1)}$ to $\xi_d \approx 1.02 \xi_d^{(1)}$. We see that the estimates
from $\g$ and $g^{(1)}$ are strongly correlated, so that in the disordered
phase nothing is gained by studying also the higher projections. 

Further investigations of the cluster-diameter distribution in the
disordered phase of the two-dimensional Ising and 3-state Potts models
revealed, however, that the new observable is not always advantageous.
Our results for the Ising model from a very long simulation of a 
$80 \times 80$ lattice with 
${\rm MCS}/\tau_{{\rm int},e} \approx 12\,288\,000 = N_{\rm meas}$
in the disordered phase at $\beta = 0.70340888 \approx 0.8 \beta_c$ are 
shown in Fig.~\ref{fig:gdiam_is_eff}. For $\beta < \beta_c$
the exact expression for the 2D Ising model correlation length is
$\xi_d = 1/(\beta^{\star} - \beta)$ \cite{ising}, where the dual inverse 
temperature
$\beta^{\star}$ is given by $(\exp(\beta)-1)(\exp(\beta^{\star}-1) = q =2$.
We see that here the $\xi_d^{\rm eff}$ derived
from $\Gd$ clearly overshoot the exact value of $\xi_d = 2.6202906\dots$
before they slowly approach it from above. Notice that $\beta$ was adjusted
such that $\xi_d$ agrees roughly with the value of the $q=20$ model at 
$\beta_t$. The $\xi_d^{\rm eff}$ of $\g$,
on the other hand, coincide with the exact value already for very small $x$,
and a simple two-parameter fit of the form (\ref{eq:fit_4}) with $b=c=0$
in the range $x=1,\dots,40 = L/2$ yields $\xi_d=2.62029(14)$, in perfect 
agreement with the exact result. A fit of $\Gd$ according to (\ref{eq:Gdfit})
using only large $x$ values in the interval $x=40,\dots,56$ gives a 
considerably higher estimate of $\xi_d = 2.89(8)$ which, despite its 
large error bar, is only barely compatible with the theoretical value. The 
deviation is still about $10\%$.

If we choose $\beta = 0.7891847 \approx 0.9 \beta_c$ (the dual
inverse temperature of $\beta^{\star}=0.98$), such that the exact 
correlation length is twice as large, $\xi_d = 5.2405812\dots$, we obtain
qualitatively the same picture. This is shown in Fig.~\ref{fig:gdiam_is160_eff}
for a simulation of a $160 \times 160$ lattice with
${\rm MCS}/\tau_{{\rm int},e} \approx 3\,200\,000 = N_{\rm meas}$. Here 
the linear fit of the data for $\g$ in
the range $x=1,\dots,80 = L/2$ yields $\xi_d = 5.2400(6)$, again in nice
agreement with the exact value. A fit of $\Gd$ in the interval 
$x=80,\dots,100$ with $\xi_d = 5.6(3)$, on the other hand, deviates again
considerably by about $7\%$.

Finally we show in Fig.~\ref{fig:gdiam_q=3_eff} our results for the
two-dimensional $q=3$ Potts model at 
$\beta = 0.95179503 \approx 0.95 \beta_c$ (where
$\beta^{\star}=1.06$). Here the lattice size is
$160 \times 160$, ${\rm MCS}/\tau_{{\rm int},e} \approx 2\,285\,000$, and
$N_{\rm meas} = 3\,200\,000$. For small $x$ we observe the influence of
higher excitations in $\g$ which, however, die out rapidly. Discarding
therefore the smallest distances and choosing a fit interval of 
$x=7,\dots,80$ we obtain an estimate of $\xi_d = 5.838(2)$, which is shown
as the horizontal line in the lower plot of Fig.~\ref{fig:gdiam_q=3_eff}.
Here the cluster-diameter distribution $\Gd$ is already slightly better behaved
than for the two-dimensional Ising model, and a fit in the interval
$x=80,\dots,106$ gives a compatible value of $\xi_d = 6.0(2)$, which 
now deviates only by $2\%$ from the result of $\g$.
%
%---------------------------------------------------------
                        \section{Discussion}
%---------------------------------------------------------
%
Our numerical results clearly show that the cluster-diameter distribution
$\Gdx$ is very well suited to extract the correlation length $\xi_d(\beta_t)$
of two-dimensional $q$-state Potts models with relatively large values of $q$.
While analyses of the standard (projected) two-point function are plaqued by
large systematic errors, with the new observable we succeeded for the first
time to reproduce the theoretically expected values at a $1\% - 2\%$ level.

For small values of $q$, however, the standard correlation function
gives much more reliable results. For reasons not well understood to date,
the two quite different correlators thus seem to behave complementary to each
other.

Also for the three-dimensional $q$-state Potts models with $q=3$, 4, and 5,
which undergo a first-order phase transition already for $q \ge 3$,
our results \cite{tobe} for $\Gd$ and $\g$ in the disordered phase at
the transition point $\beta_t$ as well as the corresponding effective 
correlation lengths look qualitatively as for the 2D Ising model
in Fig.~\ref{fig:gdiam_is_eff}. Also in these cases we found that $\g$ gives
much more reliable estimates of $\xi_d$. This suggests that the behavior
of $\Gd$ does depend crucially on the value of $q$, but certainly not on the
fact that the two-dimensional Potts models with $q=10$, 15, and 20 were
studied at their first-order transition point $\beta_t$. The details of
the 3D study will be published elsewhere \cite{tobe}.
%
%---------------------------------------------------------
                         \section*{Acknowledgements}
%---------------------------------------------------------
%
WJ thanks the DFG for a Heisenberg fellowship and 
SK gratefully acknowledges a fellowship by the
Graduierten\-kolleg ``Physik and Chemie supra\-moleku\-larer Systeme''.
Work supported by computer grants hkf001 of HLRZ J\"ulich and
bvpf03 of Norddeutscher Vektorrechnerverbund (NVV) Berlin-Hannover-Kiel.
%
%-------------------------------------------------------------------
%

\clearpage\newpage
%
%-------------------------------------------------------------------
%                             Table 1
%-------------------------------------------------------------------
%
\begin{table}[b]
\newlength{\digitwidth} \settowidth{\digitwidth}{\rm 0}
\catcode`?=\active \def?{\kern\digitwidth}
% -----------------------------------------------------
\caption[a]
   {\label{tab:stat}
   The average and maximum cluster size $\langle|C|\rangle$ and 
   $|C|_{\rm max}$, the maximum cluster diameter (diam $C_{i_0}$)$_{\rm max}$,
   the integrated autocorrelation time $\tau_{\rm int,e}$
   of the energy, the number of Monte Carlo update sweeps (MCS)
   in units of $\tau_{\rm int,e}$, and the number of measurements 
   $N_{\rm meas}$.}
\begin{center}
\begin{tabular}{|l|r|r|r|}
\hline
 \multicolumn{4}{|c|}{$\beta=\beta_t$, disordered phase, 
                      single-cluster algorithm}\\
\hline
                               &
 \multicolumn{1}{c|}{$q=10$}   &
 \multicolumn{1}{c|}{$q=15$}   &
 \multicolumn{1}{c|}{$q=20$}   \\
                               &
 \multicolumn{1}{c|}{$300\times300$}  &
 \multicolumn{1}{c|}{$120\times120$}  &
 \multicolumn{1}{c|}{$80\times80$}   \\
\hline
 $\langle|C|\rangle_{_{\rm SC}}$ &  38.093(17)  & 10.2254(11)  & 5.87776(75)   \\
 $\langle|C|\rangle_{_{\rm SW}}$ &  2.726411(61)& 2.117866(18) & 1.854135(26)  \\
 $|C|_{\rm max}$             &  9353        & 1908         & 846           \\
 (diam $C_{i_0}$)$_{\rm max}$&  ?197        & ??73         & ?51           \\
 $\tau_{{\rm int},e}$        & $\approx 59$ & $\approx 18$ & $\approx  25$ \\
 MCS/$\tau_{{\rm int},e}$    &  ?\,600\,000 & ?9\,000\,000 &  ?4\,200\,000 \\
 $N_{\rm meas}$              &  1\,600\,000 & 12\,800\,000 &  10\,395\,000 \\
\hline
\end{tabular}
\end{center}
\end{table}
%
%-------------------------------------------------------------------
%                             Table 2
%-------------------------------------------------------------------
%
\begin{table}[thb]
\begin{center}
  \caption[a]
    {\label{tab:xi.2Lx2L.fit}
    Results for $\xi_d(\beta_t)$ from two-parameter fits (\ref{eq:Gdfit})
    to $\Gdx$ for different
    fit intervals $x_{\rm min} \dots x_{\rm max}$, with
    $x_{\rm max} = 130$, 50, and 40 for $q=10$, 15, and 20, respectively.
    Also shown are results of four-parameter fits to $g^{(0)}(x)$
    and $g^{(1)}(x)$ according to the Ansatz (\ref{eq:fit_4}), (\ref{eq:xi}),
    with $x_{\rm max} = 150$, 60, and 40 for $q=10$, 15, and 20, respectively.
    }
\vspace*{0.3cm}
\begin{tabular}{|c|cl|cl|cl|}
\hline
\multicolumn{1}{|c|}{observable} &
\multicolumn{2}{c|}{$q=10$}       &
\multicolumn{2}{c|}{$q=15$}       &
\multicolumn{2}{c|}{$q=20$}       \\
\multicolumn{1}{|c|}{ } &
\multicolumn{2}{c|}{$300 \times 300$}      &
\multicolumn{2}{c|}{$120 \times 120$}       &
\multicolumn{2}{c|}{$80  \times 80$}       \\
\hline
exact & \multicolumn{2}{|r|}{$10.559519...$} & 
        \multicolumn{2}{|r|}{$4.180954...$}  &
        \multicolumn{2}{|r|}{$2.695502...$} \\
\hline
   & $x_{\rm min}$ & \multicolumn{1}{c|}{$\xi_d$} 
   & $x_{\rm min}$ & \multicolumn{1}{c|}{$\xi_d$}
   & $x_{\rm min}$ & \multicolumn{1}{c|}{$\xi_d$} \\
\hline
 $\Gd$    &40  &  10.90(2) &  20&  4.297(4) & 13&  2.766(3)   \\
          &48 &   10.92(3) &  23&  4.286(5)&  15&  2.761(3)   \\
          &56 &   10.88(3) &  26&  4.267(8)&  17&  2.752(5)   \\
          &64 &   10.84(5) &  29&  4.26(2) &  19&  2.744(7)   \\
          &72 &   10.75(8) &  32&  4.25(2) &  21&  2.74(1)   \\
          &80 &   10.6(2)  &  35&  4.27(3) &  23&  2.73(2)   \\
          &88 &   10.5(2)  &  38&  4.25(4) &  25&  2.70(3)  \\
          &96 &   10.3(3)  &  40&  4.23(6) &  27&  2.68(4)  \\
\hline
$g^{(0)}(x)$ &10 &   8.9(1)   &  5 &  3.56(2) &  3  &  2.30(1)  \\
             &12 &   9.0(1)   &  6 &  3.60(2) &  4  &  2.33(1)  \\
             &14 &   9.1(2)   &  7 &  3.63(2) &  5  &  2.36(2)  \\
             &16 &   9.2(2)   &  8 &  3.66(3) &  6  &  2.39(3)  \\
             &20 &   9.4(3)   &  9 &  3.70(4) &  7  &  2.41(4)  \\
             &22 &   9.5(4)   & 10 &  3.73(5) &  8  &  2.43(5)  \\
             &24 &   9.7(6)   & 11 &  3.76(6) &  9  &  2.46(6)  \\
%             &28 &   10.5(1.3)& 12 &  3.80(7) & 10  &  2.48(9)  \\
\hline
% These are the \xi_d as computed from \xi_d^{(1)}
%
$g^{(1)}(x)$ &10 &   8.88(7) &  5 &   3.55(1)&   3 &   2.293(7)  \\
             &12 &   8.96(9) &  6 &   3.59(2)&   4 &   2.33(1)   \\
             &14 &   9.1(1)  &  7 &   3.62(2)&   5 &   2.36(2)   \\
             &16 &   9.2(2)  &  8 &   3.65(3)&   6 &   2.38(2)   \\
             &20 &   9.3(3)  &  9 &   3.69(3)&   7 &   2.40(3)   \\
             &22 &   9.5(3)  & 10 &   3.72(4)&   8 &   2.42(4)   \\
             &24 &   9.6(4)  & 11 &   3.75(5)&   9 &   2.44(6)   \\
%             &28 &   9.8(8)  & 12 &   3.72(6)&  10 &   2.41(7)   \\
%
% The following are the directly measured \xi_d^{(1)} !!!
%
%$g^{(1)}(x)$ &10 &   8.73(7) &  5 &   3.49(1)&   3 &   2.257(7)  \\
%             &12 &   8.81(9) &  6 &   3.53(2)&   4 &   2.29(1)   \\
%             &14 &   8.9(1)  &  7 &   3.56(2)&   5 &   2.32(2)   \\
%             &16 &   9.0(2)  &  8 &   3.59(3)&   6 &   2.34(2)   \\
%             &20 &   9.1(3)  &  9 &   3.62(3)&   7 &   2.36(3)   \\
%             &22 &   9.3(3)  & 10 &   3.65(4)&   8 &   2.38(4)   \\
%             &24 &   9.4(4)  & 11 &   3.68(5)&   9 &   2.40(6)   \\
%             &28 &   9.8(8)  & 12 &   3.72(6)&  10 &   2.41(7)   \\
\hline
\end{tabular}
\end{center}
\end{table}
\clearpage
%
%-------------------------------------------------------------------------------
%                        FIGURES
%-------------------------------------------------------------------------------
%
%-------------------------------------------------------------------
%                         Figure 1
%-------------------------------------------------------------------
%
\begin{figure}[th]
\vskip 4.25truecm
\includegraphics{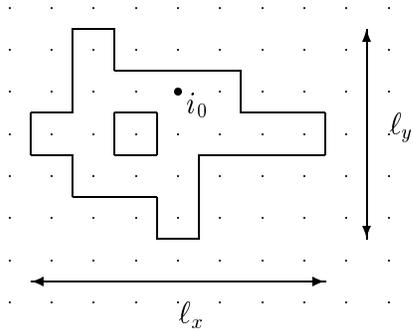}
\caption[a]{\label{fig:sketch}%
Illustration of the definition of the cluster diameter
${\rm diam} \,C_{i_0} = {\rm max}\{\ell_x,\ell_y\}$.}
\end{figure}
%
%
%-------------------------------------------------------------------
%                         Figure 2
%-------------------------------------------------------------------
%
\begin{figure}[t]
\vskip 14.00truecm
\includegraphics{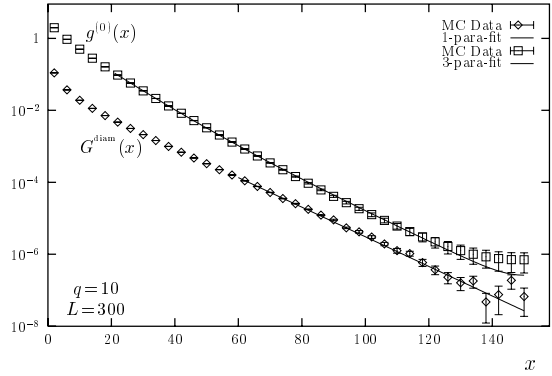}
\includegraphics{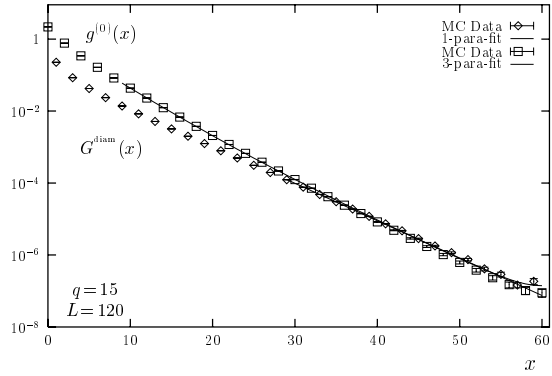}
\includegraphics{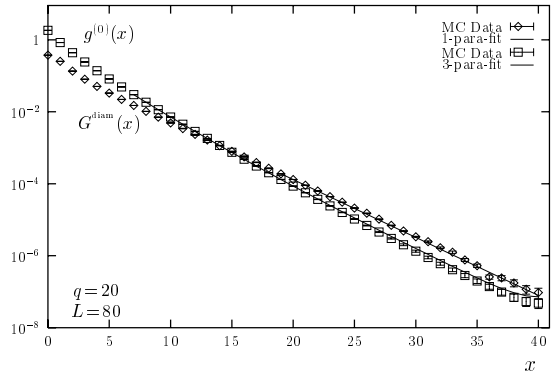}
\caption[a]{\label{fig:gdiam}%
Semi-log plot of the cluster-diameter distribution $\Gdx$ and the
projected correlation function $g^{(0)}(x)$ for $q=10$, 15, and 20 
at $\beta_t$ in the disordered phase.
For clarity some data points are omitted for $q=10$ and 15.}
\end{figure}
%
%-------------------------------------------------------------------
%
%
%-------------------------------------------------------------------
%                         Figure 3
%-------------------------------------------------------------------
%
\begin{figure}[tbh]
\vskip 14.00truecm
\includegraphics{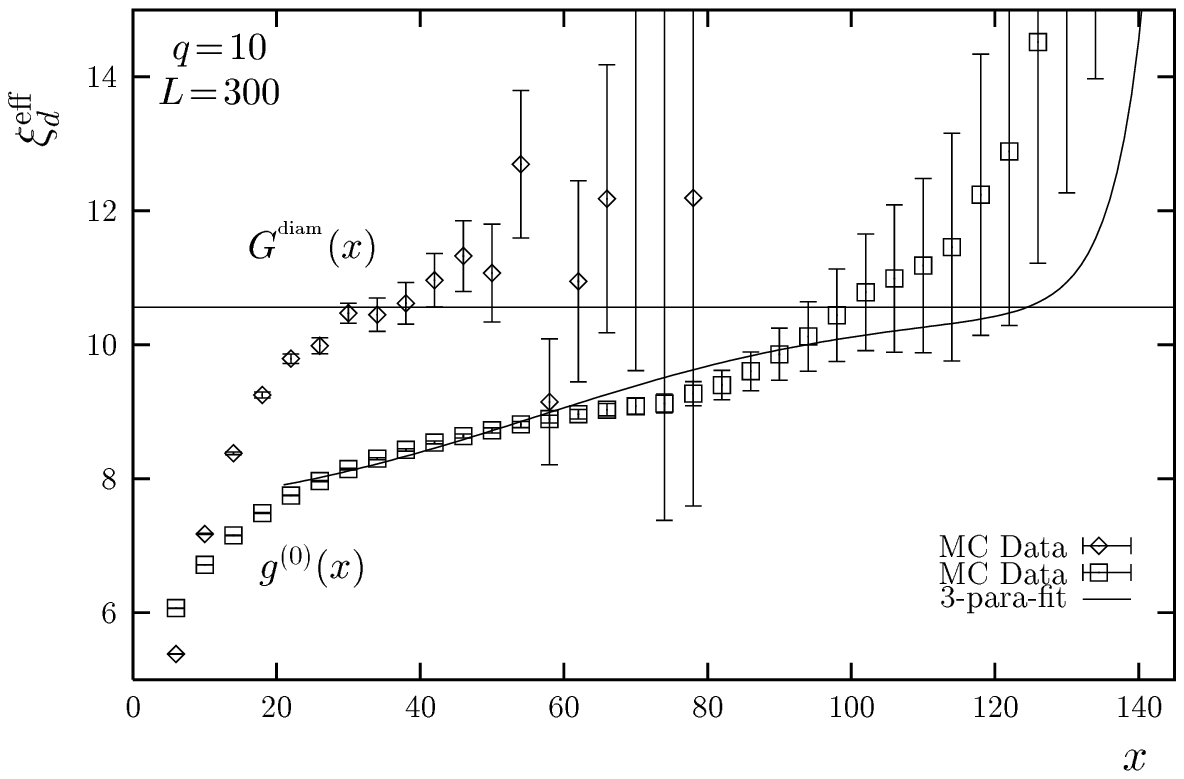}
\includegraphics{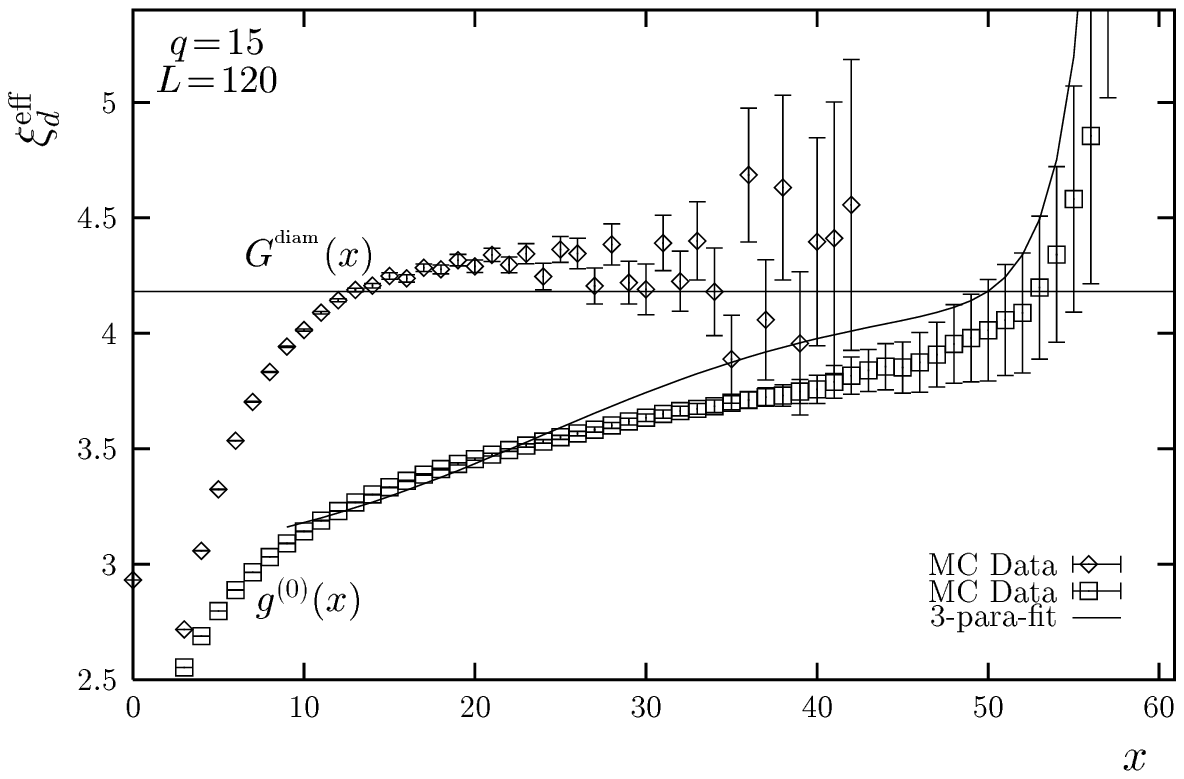}
\includegraphics{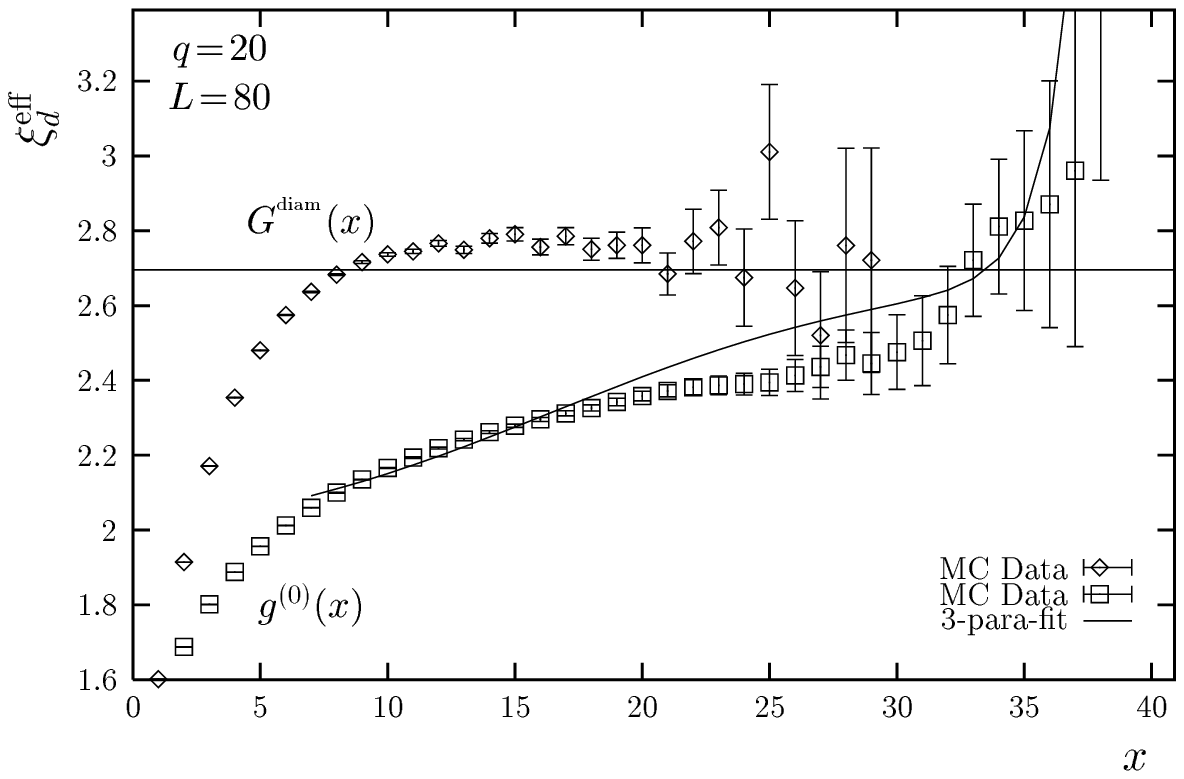}
\caption[a]{\label{fig:gdiam_eff}%
Effective correlation lengths for $q=10$, 15, and 20 at $\beta_t$ in the
disordered phase derived from the correlation functions shown in
Fig.~\ref{fig:gdiam}.}
\end{figure}
%
%-------------------------------------------------------------------
%
%
%-------------------------------------------------------------------
%                         Figure 4
%-------------------------------------------------------------------
%
\begin{figure}[tbh]
\vskip 8.0truecm
\includegraphics{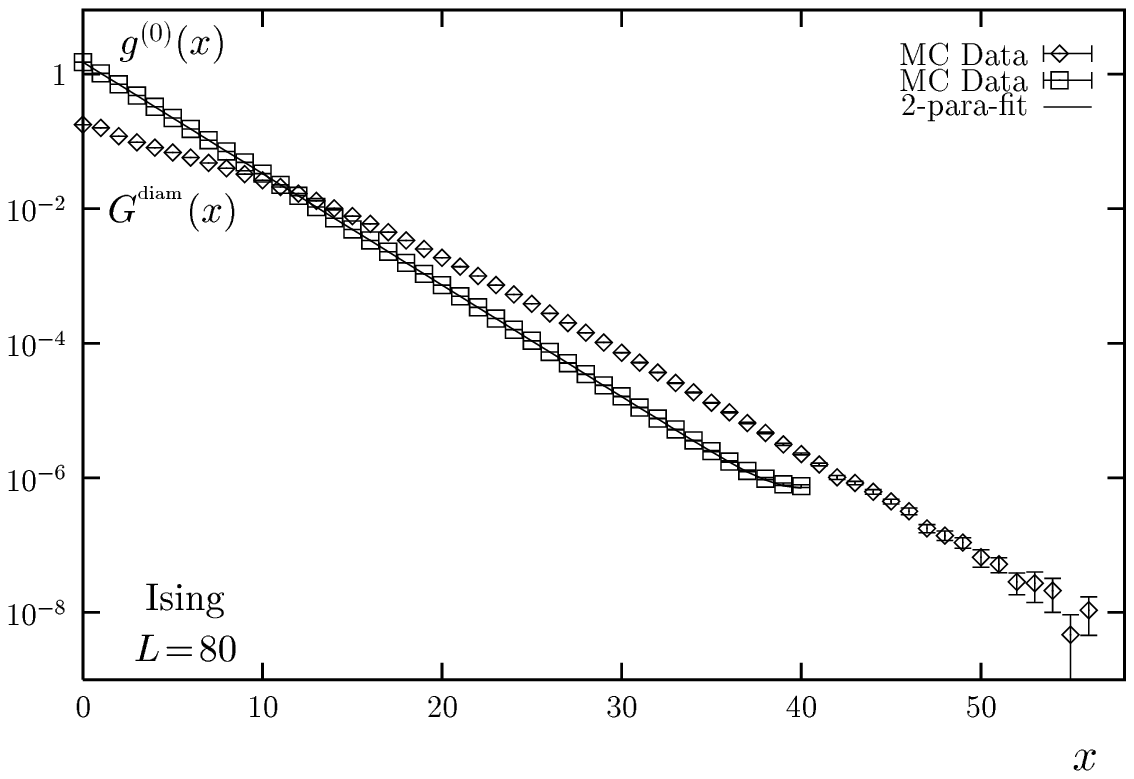}
\includegraphics{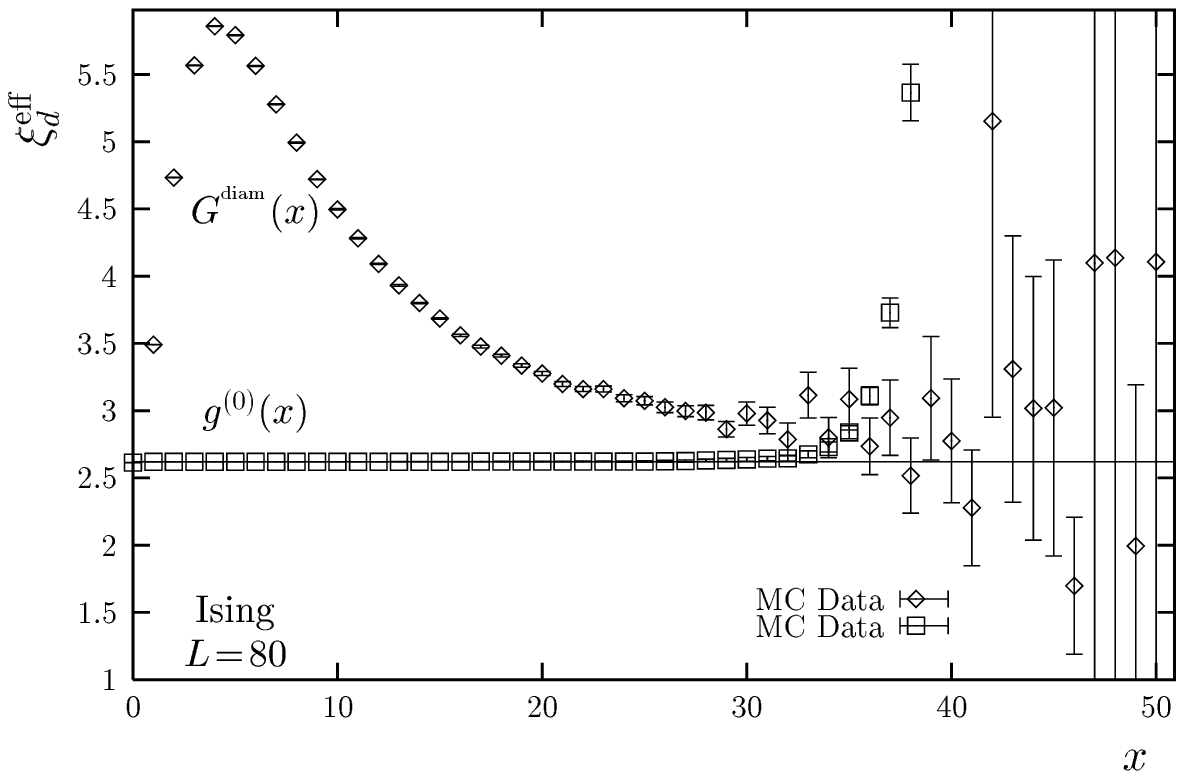}
\caption[a]{\label{fig:gdiam_is_eff}%
Correlation functions (upper plot) and
effective correlation lengths (lower plot) for the 2D Ising model 
at $\beta = 0.70340888$.
The horizontal line shows that exact value of $\xi_d = 2.6202906\dots$.}
\end{figure}
%
%-------------------------------------------------------------------
%
%
%-------------------------------------------------------------------
%                         Figure 5
%-------------------------------------------------------------------
%
\begin{figure}[tbh]
\vskip 8.0truecm
\includegraphics{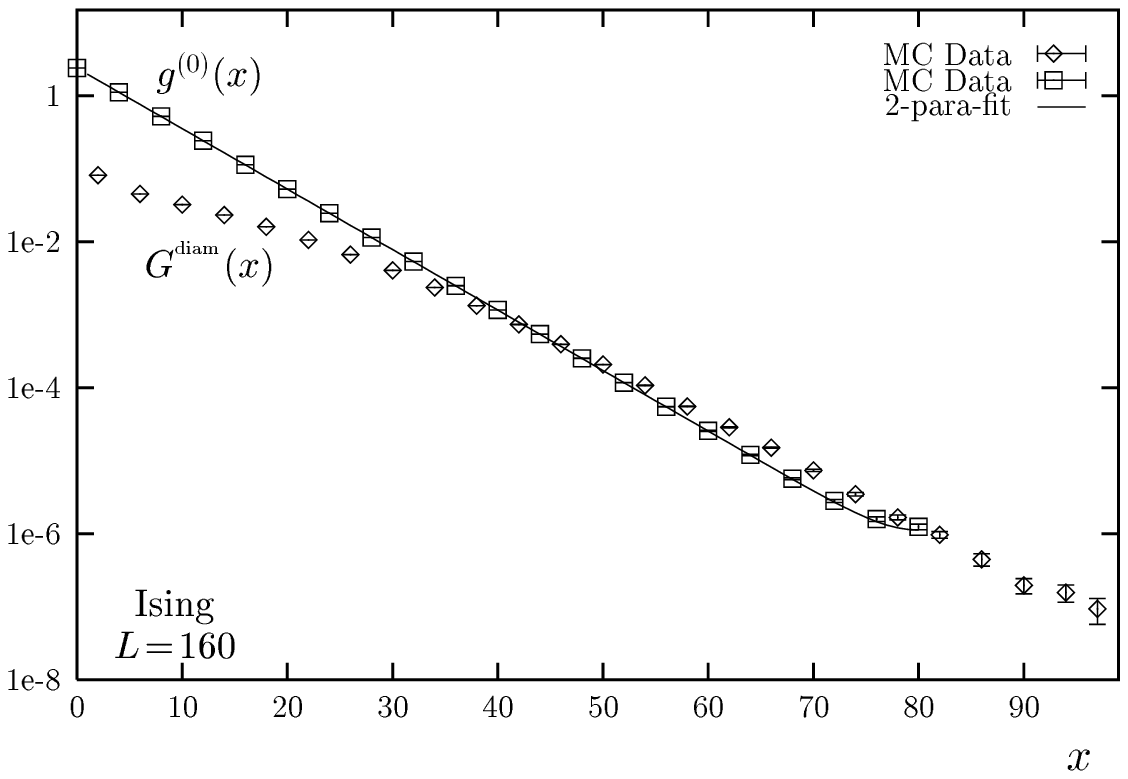}
\includegraphics{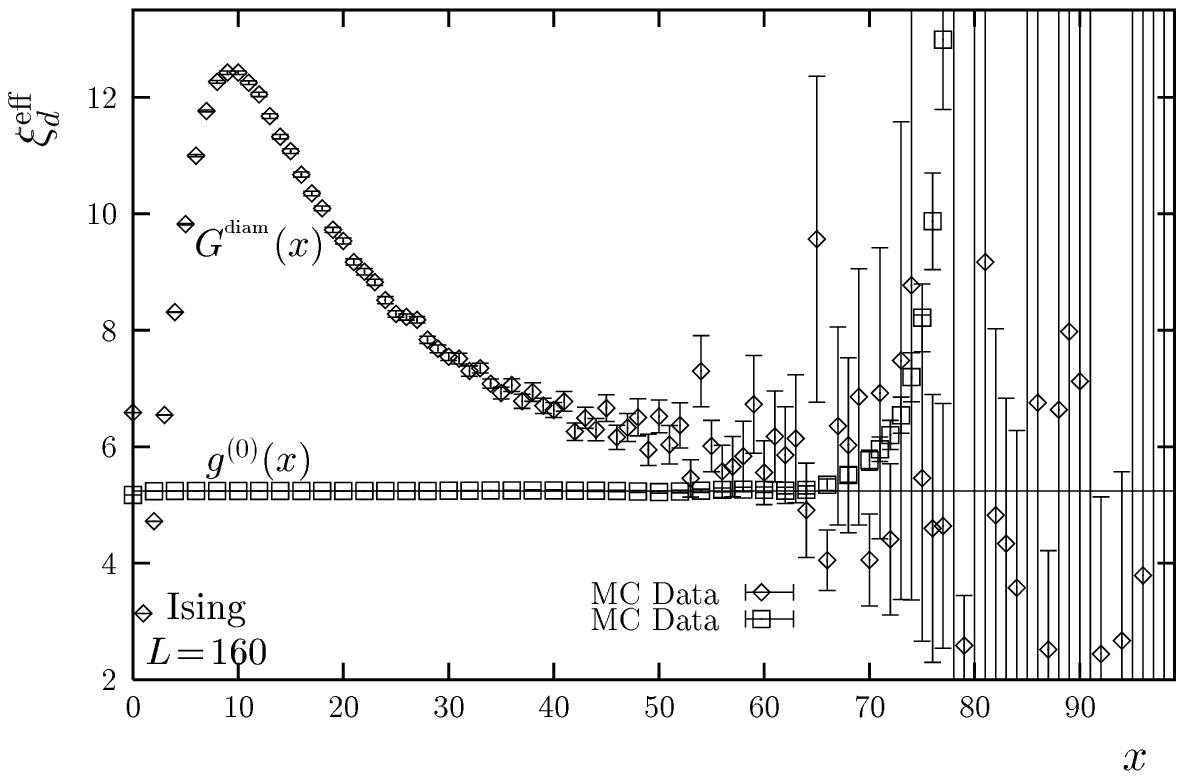}
\caption[a]{\label{fig:gdiam_is160_eff}%
Correlation functions (upper plot) and
effective correlation lengths (lower plot) for the 2D Ising model 
at $\beta = 0.78918147$.
The horizontal line shows that exact value of $\xi_d = 5.2405812\dots$.}
\end{figure}
%
%-------------------------------------------------------------------
%
%
%-------------------------------------------------------------------
%                         Figure 6
%-------------------------------------------------------------------
%
\begin{figure}[tbh]
\vskip 8.0truecm
\includegraphics{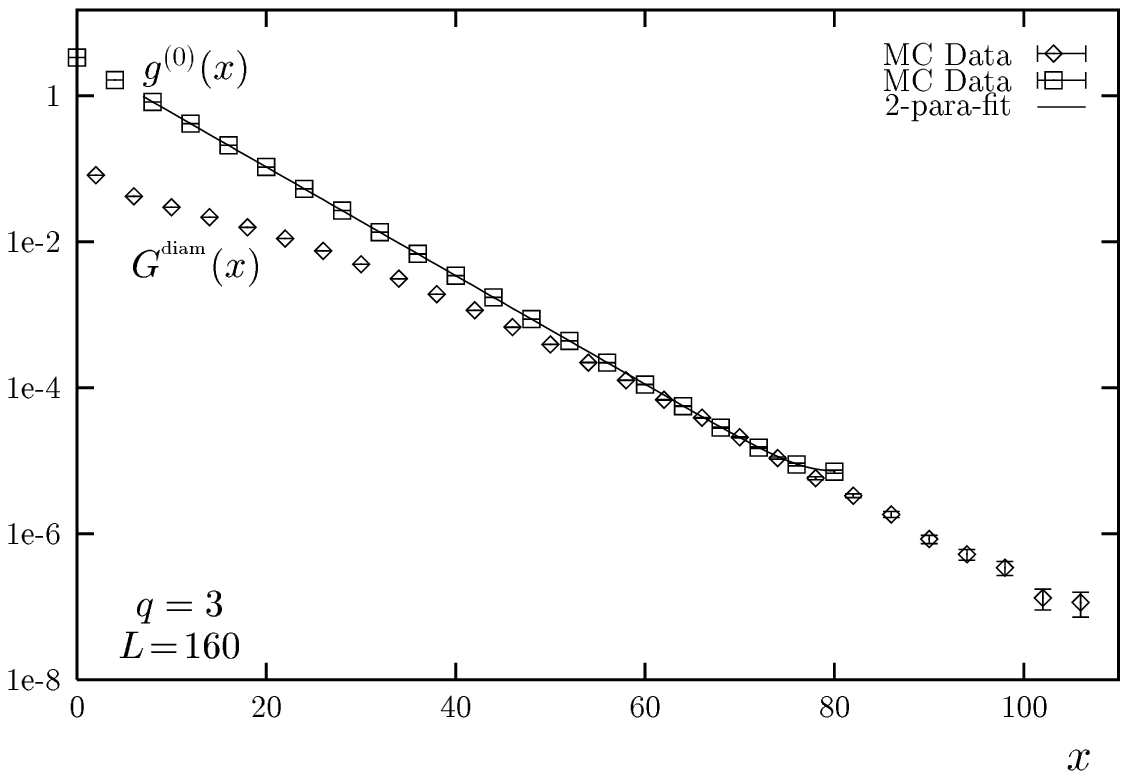}
\includegraphics{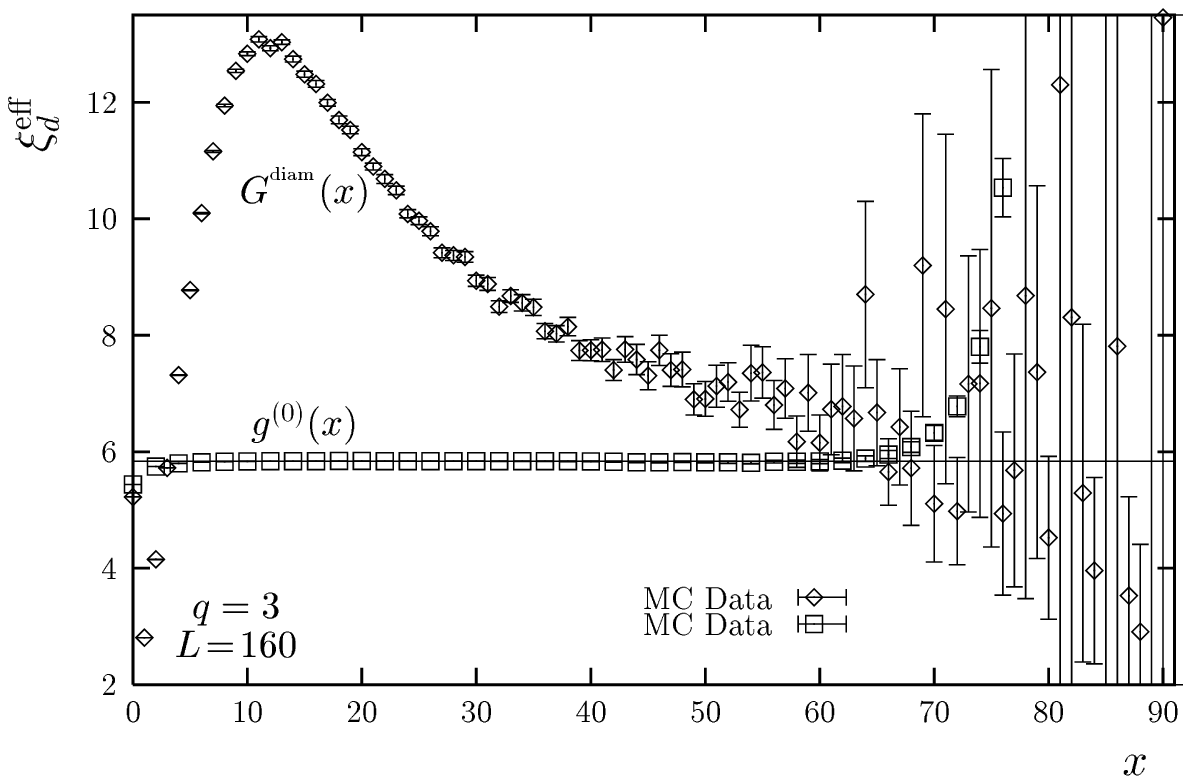}
\caption[a]{\label{fig:gdiam_q=3_eff}%
Correlation functions (upper plot) and
effective correlation lengths (lower plot) for the 2D 3-state Potts model 
at $\beta = 0.95179503$.
The horizontal line shows our best numerical estimate of $\xi_d = 5.838(2)$.}
\end{figure}
%
%-------------------------------------------------------------------
\end{document}